**Non-linear deformation and break up of enclaves in a rhyolitic magma: a case study from Lipari Island (Southern Italy)**

Francois Holtz[1], Sascha Lenné[1], Guido Ventura[2], Francesco Vetere[1], Philipp Wolf[1]

[1]Institut für Mineralogie, Universität Hannover, Hannover, Germany
[2]Istituto Nazionale di Geofisica e Vulcanologia, Roma, Italy

Corresponding author: Guido Ventura, Istituto Nazionale di Geofisica e Vulcanologia
Via di Vigna Murata 605 - 00143 Roma, Italy
Phone (+39) 06-51860221
e-mail: ventura@ingv.it

**Abstract** A dome from Lipari Island (Southern Italy) consists of 12 vol.% of circular, elongated and folded latitic enclaves hosted in a rhyolitic matrix. The dm- to cm-scale enclaves are more deformed than the mm-scale blobs. The critical value of the ratio between the viscous forces, which allow deformation and eventually break up blobs of latitic magma, and the interfacial tension forces is larger than 0.29. The Reynolds number is $\leq$ 5.3. The equivalent radius and the axial ratio of the enclaves follow power-law distributions. This feature suggests that the break up and stretching of magmas are non-linear, scale-invariant, probably cyclic processes. The coexistence of enclaves of different shape and the self-similar size distributions suggest that chaotic advection plays a major role in the formation of mingled magmas. Caution must be used when measuring the finite strain from enclave shapes because they may break apart during the deformation.

1. Introduction

The occurrence of enclaves in lava flows and domes demonstrates mingling processes between magmas of different composition. Mechanisms proposed for the formation of enclaves include [*Thomas and Tait,* 1997 and reference therein; *Coombs et al.,* 2002]: (a) breaking of a solid magma layer by injection of new magma, (b) disruption of the stratification of magma chamber during an eruption, (c) vesiculation of a mafic magma within a densely stratified reservoir, (d) injection of mafic magma into a silicic one, (e) separation and floating of enclaves from a layer of mafic magma into a more evolved magma due to coupled crystallization and vesiculation. Independently from the mechanism of formation, the shape of enclaves reflects the competition between the viscous forces, which tend to strain and potentially break apart blobs, and the interfacial tension forces, which drive blobs towards spherical shapes [*Williams and Tobish*, 1994]. These two forces act before the formation of a chilled margin at the enclave/host interface [*Blake and Fink,* 2000]. Studies on enclaves are relatively abundant [see *Coombs et al.*, 2002 and *Paterson et al.*, 2004 for a review], however, little is known about the mechanisms of deformation and break up as well as about the dynamics of mingling process. Fluid-dynamical experiments on mingling of immiscible viscous liquids show that three main processes occur during the flow of a dispersed phase hosted in a continuous phase [*Ottino et al.*, 2000]: stretching (stirring), folding and break up. These three processes may act simultaneously in different flow types (e.g. shear, hyperbolic and extensional flows) and at length scales.

In this study, we analyze the latitic enclaves hosted in a rhyolitic lava dome of Lipari Island (Fig. 1a; Aeolian Islands, Southern Tyrrhenian Sea). We determine the main physical parameters of the dome (strain rate, time of emplacement, and viscosities of the mingled magmas) and calculate the ratio between the viscous forces and the interfacial tension forces needed to deform and eventually break up the latitic magma throughout dimensional analysis of the enclaves. We also calculate the Reynolds number and show that the break up and stretching are scale-invariant processes that may occur in a laminar flow regime. The collected data are discussed in light of results from fluid-dynamical models and allow us to conclude that chaotic advection (fractal stirring, folding and break up) plays a major role in the formation of mingled magmas.

## 2. Geological setting and physical volcanology of the Lipari dome

Subaerial volcanism at Lipari (Aeolian Islands, Fig. 1a) dates back to about 220 ka [*Crisci et al.,* 1991] and historical reports indicate the date of 580 A.D. for the latest volcanic activity. S. Nicola dome 2 (hereafter SND2 in Fig. 1a), the dome studied here, crops out in the southeastern sector of Lipari and it is named. It represents the central extrusion of three domes aligned NNW-SSE and erupted between 16.8 and 23.5 ka. SND2 is an axisymmetric-type exogenous dome and measures 43 m in height and about 210 m in diameter. The poorly exposed substratum is subhorizontal and consists of older lavas. SND2 external portions contain blocks 0.5 to 3 m in diameter, whereas the interior is massive and texturally heterogeneous (Fig. 1b). Dark enclaves are latitic in composition [$SiO_2$= 59-63 wt.%; $K_2O$= 4-4.4 wt.%; *Gioncada et al.,* 2003] with 25 vol.% of crystals (clinopyroxene, plagioclase, olivine, biotite) are hosted in a light grey, rhyolitic matrix [$SiO_2$= 74-75.5 wt.%; $K_2O$= 4.5-5 wt.%] with 0-4 vol.% of crystals (oligoclase+K-feldspar+hornblende). The commingled rhyolite and latite of the dome resulted from the intrusion of a latitic dike into a shallower rhyolitic reservoir [*Gioncada et al.,* 2003].

A rough estimate of the SND2 emplacement time has been calculated using the empirical relation between dome height *h* and emplacement time *t* reported by *Fink and Griffiths* [1998, see their fig. 10] and a value of *t* of 5 days has been obtained. The volume *v* of the dome is $8.4 \cdot 10^5$ m$^3$, and the calculated effusion rate *E=v/t* is 1.93 m$^3$/s. The strain rate $\gamma$=*E/v* is $2.3 \cdot 10^{-6}$ s$^{-1}$.

### 3. The SND2 latitic enclaves

The SND2 enclaves have been observed in a well-exposed, vertical N-S section (Fig. 1a) that contains the transport direction and crosses the medium-upper zone of the dome. Here, the interface between enclaves and the rhyolitic host is well defined. A rim of small-diameter (<0.3 cm) vesicles sometimes surrounds the enclaves and documents, according to *Blake and Fink* [2000], the local occurrence of freezing.

At a dm- to cm-scale, enclaves show four main shapes (Fig. 1b): subcircular (11 %), elongated (stretched) (85 %), folded (3 %), and angular (1%). At a smaller, cm- to mm- scale, only circular (73 %) and stretched shapes (27 %) occur. The stretched and poorly folded shapes suggest that the enclaves deformed during the mingling within their host, whereas the angular shape indicates that a low percentage (1%) of the enclaves behaved as a solid.

The number of enclaves outcropping in the N-S vertical section of SND2 has been measured in an area of 30.1 m$^2$ by field mapping (enclaves with linear dimension > 1 cm) and under the microscope (enclaves with linear dimension $\leq$ 1 cm) using 4.5 x 3.5 cm thin sections. We select this N-S vertical section because it contains the flow direction and it is orthogonal to the substratum on which the flow moved. Then, the aspect ratio of the enclaves measured in this outcrop is representative of the maximum deformation, as also shown by *Ventura* [2004] on the basis of structural data.

In the selected outcrop, 170 thin sections were collected at random position and cut from rock samples parallel to the exposed surface. The area covered by the thin sections is 0.26 m$^2$. Because the area covered by the enclaves with linear size larger than 1 cm is 1.1 m$^2$, the resulting area to be investigated with thin sections should be 29 m$^2$, i.e. the difference between 30.1 and 1.1 m$^2$. In order to cover the whole surface of the outcrop, the number of enclaves counted in the 170 thin sections has been multiplied by 111 (=29 m$^2$/0.26 m$^2$). This latter procedure is justified by the observation that the difference in the number of counted enclaves in the different thin sections is less than 4 vol.%.

Following the above described procedure, we estimate that the enclaves represent 12 % of the exposed surface. The parameters determined for each stretched or circular enclave are: the axial ratio $R_f$ =long axis/short axis, the perimeter *P*, the area *A*, the radius *r* of the circle of equal area. These parameters have been determined using the image analysis ScionImage software by Scion Corporation.

## 4. Results

Results of the measurements carried out on the SND2 enclaves indicate that the distribution of the parameter $r$ (Fig. 2a) follows a power-law $N_i=C/r_i^D$, where $N_i$ is the number of objects with a characteristic dimension $r_i$, $C$ is a constant of proportionality and $D$ is the fractal dimension. As a result, the distribution of $r$ is scale-invariant ($D=1.82\pm0.19$) within three orders of length scale (0.1-10 cm). $A$ and $P$ of enclaves (Fig. 2b) are correlated by $P\sim\sqrt{A}^D$ with $D=1.03\pm0.10$, which is consistent with a fractal fragmentation process [*Lovejoy,* 1982; *Turcotte,* 1989]. The power-law distribution of $r$ and the relationships between $A$ and $P$ indicate that that the break-up (fragmentation) of the SDN2 enclaves is a scale-invariant, non-linear process. The low value of $D$ in the *P-A* relation reflects the high proportion of sub-circular enclaves. The distribution of $R_f$ (Fig. 2c), which is a measure of the stretching, also follows a power-law with $D=2.66\pm0.25$. In a $r$ vs. $R_f$ plot (Fig. 2d) there is a direct relationship between these two parameters. This suggests that stretching is proportional to the size of enclaves.

## 5. Discussion and conclusions

The collected data indicate that (1) the SND2 latitic enclaves stretch and break apart during the flow within the rhyolitic host, (2) break-up and stretching are scale-invariant processes, and (3) the larger enclaves deform more than the smaller ones.

According to drop deformation theory [*Rallison*, 1984], the parameters that control the stretch of an enclave are: (1) the ratio between the viscosity of the enclave and the viscosity of the host, and (2) the capillary number $Ca=\eta a\gamma/\sigma$, where $\eta$ is the viscosity of the host, $\gamma$ is the strain rate of the host, $a$ is the radius of the initially spherical drop, and $\sigma$ is the surface tension. <u>Ca represents the ratio between the viscous forces and the interfacial tension forces.</u> The role played by these parameters in the deformation of enclaves in magmatic flows has been discussed in detail by *Williams and Tobish* [1994] and *Ventura* [2001]. In these flows, enclaves deform if (a) the ratio between the viscosity of the enclave and the viscosity of the host is less than 4, and (b) $Ca>Ca_{crit.}$, where $Ca_{crit.}$ is the critical value of $Ca$ over which continuous stretching and eventually break up occur [*DeRoussel et al.*, 2001 and references therein].

At SND2, the viscosity of the latitic and of the rhyolitic glasses has been determined following *Shaw* [1972] for the latite at 1080°C and $H_2O$=2 wt.%, and *Hess and Dingwell* (1996) for the rhyolite at 760°C and $H_2O$=5 wt%. The effect of crystals has been also included following *Pinkerton and Stevenson* [1992]. Chemical analyses, temperature, water and crystal contents determined for SND2 have been taken from *Gioncada et al.* [2003]. $H_2O$ content was constrained on the basis of hornblende in equilibrium with the rhyolitic glass at 760°C [*Gioncada et al.* 2003]. The calculated viscosities of the SND2 latitic and rhyolitic magma are $2.1 \cdot 10^2$ Pa s and $3.2 \cdot 10^5$ Pa s, respectively. Then, at the beginning of the mingling process, the ratio between the viscosity of the latitic enclaves and the viscosity of the rhyolitic host was $6 \cdot 10^{-2}$. Following the model of *Williams and Tobish* [1994], this ratio increased up to 4 when the temperature of magmas was 920°C. We conclude that the SND2 enclaves deformed within a 1028°-920°C temperature range. Because of the low percentage (12 vol.%) of the latitic magma hosted in the rhyolite, thermal exchange between the two magmas does not occur and only the cooling of the latite probably occurred, as suggested by experimental models on magma mixing [*Blake and Fink*, 2000]. In conclusion, the SND2 enclaves do not record the entire deformation history of the dome.

As concerns the estimate of $Ca_{crit}$, assuming (a) a previously calculated value of $\gamma= 2.3 \cdot 10^{-6} s^{-1}$, which is within the range of the values estimated for flow processes in silicic magmas [$10^{-7} s^{-1}<\gamma<10^{-5} s^{-1}$; *Spera et al.,* 1988], (b) $\sigma= 8 \cdot 10^{-3}$ N/m, which is the value based on field and theoretical estimates adopted by *William and Tobish,* [1994] for mingled rocks with compositions similar to the SND2 magmas, (c) $\eta = 3.2 \cdot 10^5$ Pa s, and placing $a=r$, we estimate $Ca$ for each measured enclave. The obtained $Ca$-values are between 16.5 (largest enclave, $r$ =18 cm) and 0.018 (smallest enclave, $r$= 0.02 cm). Because of (1) the direct relationships between $r$ and $R_f$ (Fig. 2d), and (2) the observation that the enclaves with $r>0.28$ cm have $R_f \geq 1$, whereas those with $r \leq 0.28$ cm have $R_f =1$, we conclude that $Ca_{crit}$ is the value of $Ca$ of the enclaves with $r=0.28$ cm, i.e. $Ca_{crit}=0.29$. This rough estimate of $Ca_{crit}$, which, on the basis of the above reported range of $\gamma$-values for silicic magmas, may vary within two order of magnitude ($0.029 < Ca_{crit} < 2.9$), has the significance of a minimum estimate because the viscosity of the host increases as the temperature of SND2 decreases during the emplacement. Nonetheless, the value of $Ca_{crit}$ determined in this way agrees well with the experimental results of *Bentley and Leal* [1986] and *Chen et al.* [2004], who determined $Ca_{crit}$ between 0.2 and 0.4 for liquids with viscosity ratio in the order of $10^{-3}$-$10^{-2}$ in two-dimensional flows of different vorticity.

In order to check if an inertia-induced shift of the $Ca_{crit}$ [*Wagner et al.,* 2003] occurred, we also calculate the Reynolds number $Re=\rho \gamma a^2/\eta$, where $\rho$ is the density of the host and the other parameters have been previously defined. Using $\rho=2300$ kg/m$^3$ [*De Rosa et al.,* 2003], we obtain $Re$-values between 5.3 (largest enclave) and $10^{-3}$ (smallest enclaves). These values suggest that inertial forces did not play a significant role in the stretching and deformation of enclaves, which according to [*Raynal and Gence,* 1995], may form at very low $Re$.

The coexistence in the same outcrop of SND2 enclaves of circular, stretched, and fold shapes (Fig. 1b) suggest that advection (stirring, folding and break-up) occurred during the magmatic flow in the conduit and on the surface. Our data also indicate that deformation and fragmentation are non-linear, self-similar processes. Results from experimental and numerical models on mingling of viscous fluids in laminar chaotic flows have shown that the distributions of stretch and of the radius of drops produced by break up are self-similar [*Janssen and Meijer,* 1993; *Ottino et al.,* 2000]. As a result, we propose that the SND2 mingling structures were produced by chaotic advection. Because of the chaotic dynamics, i.e. the dependence of the dynamical system from the initial conditions, we cannot define the degree of evolution of the mixing process. On the other hand, the prevalence of stretched enclaves (85 vol.%) at a dm- to cm- scale suggests that more potential break up of enclaves was possible before freezing. Then, the mingling process at SND2 was to initial-intermediate stages. In chaotic flows, the break up of blobs may be a repetitive processes. If the break up produces small drops with $Ca<Ca_{crit.}$, further break up of these drops does not occur. On the contrary, if the break up produce drops with $Ca>Ca_{crit.}$, more break up is allowed and mixing between magmas may occur provided that $\sigma\sim 0$ [*Williams and Tobish,* 1994], which is not the SND2 case. This observation implies that caution must be used when measuring the finite strain from enclave shapes because they may break apart during the deformation and do not necessarily form at the same time, as also suggested by *Paterson et al.* [2004] for microgranitoid enclaves on the basis of strain data.

The conclusions of this work may be summarized in four main points: (1) size distribution of stretched and sub-circular enclaves allows the estimation of $Ca$ and $Re$ of heterogeneous, mingled magmas; (2) magma mingling is a non-linear, scale invariant process; (3) in magmatic flows, chaotic advection may occur not only in a fully turbulent regime [*Perugini et al.,* 2004], but also in a laminar regime; (4) enclave shapes should be used with caution for measurements of finite strain.

**Acknowledgements**

G.V. thanks A.Gioncada, R. De Rosa and R.Mazzuoli for the information on the Lipari geology. Work founded by the MIUR project "Fluid dynamic regime of magma mixing/mingling processes" and Italy-Germany 'Vigoni' project.

**Figure Captions**

**Fig. 1.** (a) Location and simplified geological map of Lipari Island and SND2 dome (from Crisci et al. [1991] and Gincada et al. [2003], modified). (b) Photos showing the different shapes of the SND2 latitic enclaves. Arrows evidence the smaller enclaves.

**Fig. 2.** (a) $r$ vs. $N(r)$ plot of the analyzed SND2 enclaves; measurements are from the outcrop for $r>1$ cm and from thin section for $r \leq 1$ cm. (b) Perimeter ($P$) vs. surface area ($A$) of the SND2 enclaves. (c) $R_f$ vs. $N(R_f)$ plot of the SND2 enclaves. In (a), (b) and (c) plots, $D$ is the calculated fractal dimension. (d) $R_f$ vs. $r$ plot of the SND2 enclaves. The inset shows the magnification of the region of the plot where $1 \leq R_f \leq 1.4$ and $0 \leq r \leq 0.8$ cm. Note that enclaves with $r \leq 0.28$ cm have $R_f = 1$.

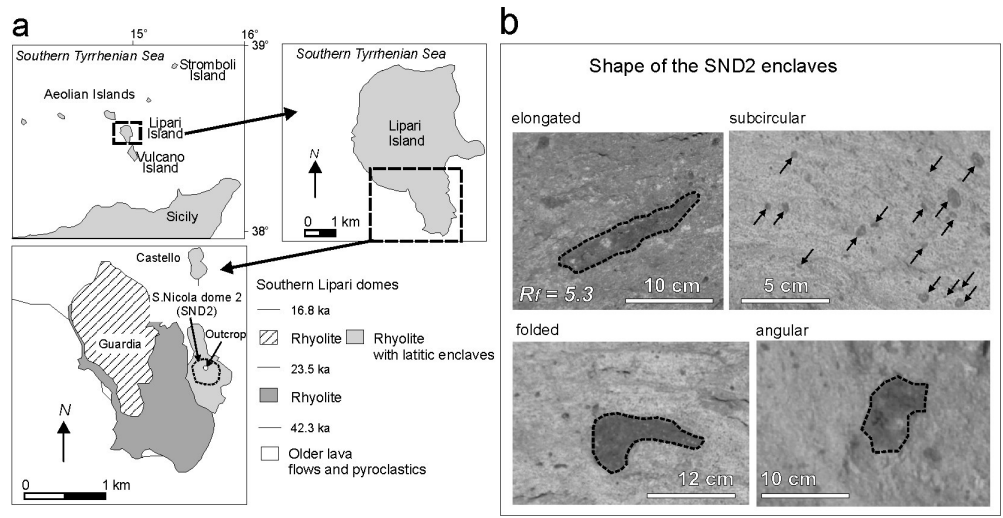

**FIG.1**

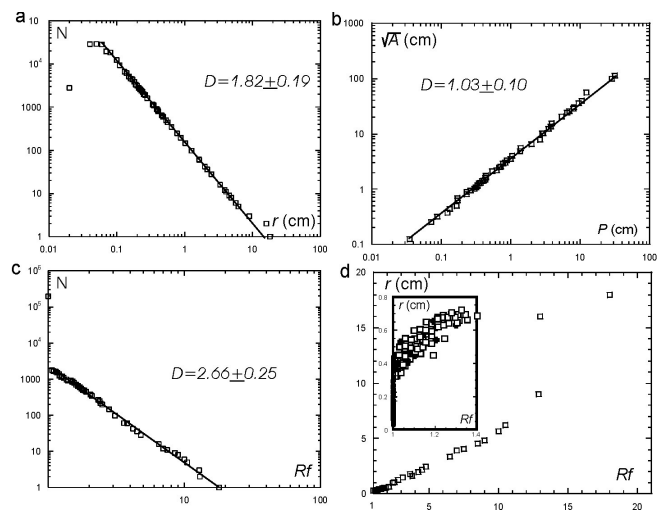

**FIG. 2**